\documentclass[aps,prd,reprint,showpacs,groupedaddress]{revtex4-1}

\usepackage{graphicx}% Include figure files

\begin{document}
 
\title{Fine-tuning in scalar meson decay with bound-state corrections}
 
\author{M. L. L. da Silva}
\email[]{mllsilva@gmail.com}
\affiliation{Instituto de F\'{\i}sica e Matem\'atica, Universidade Federal de
Pelotas, Caixa Postal 354, CEP 96010-090, Pelotas, RS, Brazil.}

\author{D. T. da Silva}
\affiliation{Instituto de F\'{\i}sica e Matem\'atica, Universidade Federal de
Pelotas, Caixa Postal 354, CEP 96010-090, Pelotas, RS, Brazil.}

\author{C. A. Z. Vasconcellos}
\affiliation{Instituto de F\'{\i}sica, Universidade Federal do Rio Grande do  Sul,
Caixa Postal 15051, CEP 91501-970, Porto Alegre, RS, Brazil}

\author{D. Hadjimichef}
\affiliation{Instituto de F\'{\i}sica, Universidade Federal do Rio Grande do  Sul,
Caixa Postal 15051, CEP 91501-970, Porto Alegre, RS, Brazil}

\date{\today}

\begin{abstract}

The glueball-quarkonia mixture decay amplitude has been evaluated, in the literature, by diagrammatic
techniques for drawing quark lines and the quark-gluon vertex. In this paper we use an alternative
approach which consists in a mapping technique, the Fock-Tani formalism, in order to obtain an
effective Hamiltonian starting from  microscopic interactions. An extra effect is present in this
formalism associated to the extended nature of mesons: bound-state corrections, which introduces an
additional decay amplitude and sets a fine-tuning procedure for general meson decay calculations. The
$f_0(1500) \to \pi \pi$ channel shall be considered as numerical example of the procedure.

\end{abstract}


\maketitle

\section{Introduction} 

The gluon self-coupling in QCD opens the possibility of existing bound states of pure gauge fields
known as glueballs. Glueballs are predicted by many models and by lattice calculations. Many
mesons have stood up as good candidates for the lightest glueball in the spectrum and in particular
the scalar sector seems promising. In the scalar sector it is expected the mixing of glueball states
with nearby quark-antiquark states (for a review see \cite{mathieu,curtis-meyer-09,ochs}). Another
motivation is the new generation of experiments that will focus on the search for exotic states. 
In particular we expect the PANDA experiment at FAIR to establish the existence of glueballs \cite{panda}.
We also expect that the GlueX and CLAS12 experiments can help us determine the structure of scalar
resonances and enable one to obtain conclusive evidence of the existence of glueballs.

The mass of low-lying scalar glueball is expected by lattice calculation to be in the range of 
$1.5 - 1.8$ GeV \cite{bali,hchen,morningstar,vaccarino,loan,ychen,gregory} and recently the glueball
spectrum and the radiative decay of $J/\Psi$ from unquenched lattice QDC and experimental data was
considered to study the scalar glueball in Ref. \cite{cheng_hep}. The most feasible glueball candidates
are scalars $f_{0}(1370)$, $f_{0}(1500)$ and $f_{0}(1710)$ \cite{pdg}. 
%Even the $f_{0}(1370)$ is out of the glueball mass range it has a small glueball content. 
The scalar sector has been studied for a long time,
but there is no agreement about the glueball content of $f_{0}(1370)$, $f_{0}(1500)$ and
$f_{0}(1710)$ \cite{amsler1,amsler2,weingarten,faessler,close1}.

In the lowest order, it is expected that a mixture of the scalar glueball $G$ and quarkonia states
$n\bar{n}=(u\bar{u}+ d\bar{d})/\sqrt{2} $ and $s\bar{s}$ should exist. There are three eigenstates $|f_0(M)\rangle$ with
physical masses $M$ given by  
\begin{eqnarray}
|f_0(M)\rangle=c_1 \,|n\bar{n}\rangle +c_2 \,|s\bar{s} \rangle + c_3 \,|G\rangle.
\label{f0}
\end{eqnarray}
with the following normalization condition $\sum_{i=1}^{3} c_{i}^{2}=1$. In the literature these
parameters have been adjusted to those of the observed resonances $f_0(1370)$, $f_0(1500)$ and
$f_0(1710)$ \cite{close1,close2,close3}.

Recently this matter has been discussed in several papers where models are used to investigate the
structure of these scalars. For example, scalar meson photoproduction was calculated at GlueX
energies \cite{magno}, these  results  could provide novel tests for our understanding
of the nature of the scalar resonances. In another calculation, exclusive production of glueballs at high energies
was performed for the scalar sector \cite{magno2}, 
where sufficiently large  cross sections were found, feasible for  experimental measurement. 
%
%and it was found cross sections which are sufficiently large for experimental measurement. 
% 
In Ref. \cite{fariborz} the authors used an effective
nonlinear chiral Lagrangian to study the $f_0(1370)$ structure. In their results they obtained that
$f_0(1370)$ was predominantly quark-antiquark state with substantial $s\bar{s}$ content. 
Resonances $f_{0}(1370)$, $f_{0}(1500)$ and $f_{0}(1710)$ were also studied in a Linear Sigma Model
\cite{giacosa1,giacosa2}, where it was shown that 
%the authors have obtained that 
$f_0(1370)$ was predominantly
quark-antiquark state, $f_{0}(1500)$ was predominantly $s\bar{s}$ state and  $f_{0}(1710)$ is a
glueball. The authors in Ref. \cite{denis} calculated the decay rates for scalar glueballs in a
Witten-Sakai-Sujimoto model and suggest the $f_{0}(1710)$ as the scalar glueball. Monte Carlo
simulation and phenomenological studies were used to study glueballs in Ref. \cite{dobbs}, in the
scalar sector their results shown that $f_{0}(1710)$ is not a pure scalar glueball. A
phenomenological study of $B$ semi-inclusive decay were performed in Ref. \cite{he} where they
found that $f_{0}(1710)$ is mainly scalar glueball.

In the experimental point of view there are uncertainties in the scalar sector for example the
$f_{0}(1710)$, first discovered by Crystal-Ball in radiative $J/\psi$  decays into $\eta\eta$
was consistent with a dominant $s\bar{s}$ assignment and confirmed by WA102 which reported a much
stronger coupling to $K \bar{K}$ than to $\pi\pi$ which spoiling the pure glueball picture. But
the BES Collaboration suggests the existence of an other resonance with mass around $1790$ MeV which
has a strong $f_0(1790) \to \pi\pi$ decay but no corresponding signal for decays to $K\bar{K}$
\cite{bes}. Another problem is low data statistics in processes which provide important results
that can be compared with theoretical models that could provide crucial information about the constituent scalar
content.

In this work we will apply a mapping technique known as the Fock-Tani formalism to describe
the strong decay of the scalar resonances. This formalism has been developed in hadron physics to
deal with scattering of composite particles  with constituent interchange \cite{plb96,annals} and
has been extended recently to composite meson decay \cite{prd08}. The novel feature  was the
presence of bound-state corrections in the decay amplitude. In the present we shall evaluate the
effect of the bound-state correction, originated from the Fock-Tani formalism, in a quark-glue
mixture for a state as the one defined in Eq.~(\ref{f0}). As a simple numerical example we shall
the impact of this fine-tuning in a particular decay channel, $f_0(1500)\to 2\pi$, shall be
considered. 

\section{ Fock-Tani formalism for Mixtures} 

In the Fock-Tani formalism, the starting point is the definition of the composite meson creation
operator $F^{\dag}$. This operator is written in a second quantization notation, but differently than
in references \cite{plb96,annals,prd08} a simple definition in terms of the elementary constituents
is not possible. The presence of a quark-gluon mixture implies in a two step procedure in order to
establish the precise meaning of the composite particles present in the theory. First, an operator
that creates quark-antiquark ${\cal Q}\,\bar{\cal Q}$ bound-state meson can written as
\begin{eqnarray}
M^{\dagger}_{\alpha}&=& \Phi_{\alpha}^{\mu \nu}
{\cal Q}^{\,\dag}_\mu\,\bar{{\cal Q}}^{\,\dag}_\nu
\label{m-meson}
\end{eqnarray}
$\Phi$ is the bound-state wave-function and ${\cal Q}^{\dag}(\bar{{\cal Q}}^{\dag})$ is the $u$, $d$, $s$
quark (antiquark) creation operator. In the compact notation, where sum (integration) is implied 
over repeated indexes, ($\mu,\nu$) are the set of momentum, spin, color, flavor for quark and
antiquarks; the $\alpha$ index is the meson's set of quantum numbers. In an explicit form, the
quarkonia and strangeonia components of Eq.~(\ref{m-meson}) can be separated as follows
\begin{eqnarray}
M^{\dagger}_{\alpha}&=& \Phi_{\alpha}^{\mu \nu}
{\cal Q}^{\,\dag}_\mu\,\bar{{\cal Q}}^{\,\dag}_\nu
=a_1\,\,\varphi_{\alpha}^{\mu \nu}
\, q_{\mu  }^{\dagger}\bar{q}_{\nu}^{\dagger}
+a_2\,\,\Upsilon_{\alpha}^{\mu \nu}
\, Q_{\mu  }^{\dagger}\bar{Q}_{\nu}^{\dagger}
\nonumber\\
&\equiv&a_1\,N^{\dagger}_{\alpha}
+a_2\,S^{\dagger}_{\alpha}
\label{m-meson2}
\end{eqnarray}
where $c_1$ and $c_2$ are the quark sector mixing parameters. The next step is the definition of the
glueball creation operator, written as a two-gluon bound-state 
\begin{eqnarray}
G_{\alpha}^{\dagger}&=&\frac{1}{\sqrt{2}}\Psi_{\alpha}^{\mu \nu}
  a_{\mu}^{\dagger}a_{\nu}^{\dagger},
\label{g-meson}
\end{eqnarray}
$\Psi$ is the bound-state wave-function and $a^{\dag}$ is the gluon creation operator. It is clear
from this procedure  that $M_{\alpha}^{\dagger}$ holds the fermionic components, which is separated
from a purely bosonic constituents in $G_{\alpha}^{\dagger}$. The last step, is the definition of
$F^{\dag}$, consistent with Eqs.~(\ref{f0})-(\ref{g-meson}). The simplest choice is a linear combination
of operators $M^\dag$ and $G^\dag$
\begin{eqnarray}
F^{\dag}_{\alpha}=b_1\,M_{\alpha}^{\dagger} + b_2\,G_{\alpha}^{\dagger}\,
\label{qq-gg}
\end{eqnarray}

The gluon, quark and antiquark   operators in the former equations satisfy the following canonical
relations,
\begin{eqnarray}
&&
\lbrack a_{\mu},a^{\dag}_{\nu}\rbrack= \delta_{\mu \nu} 
\,\,\,\,\,;\,\,\,\,\, 
\{q_{\mu}, q^{\dagger}_{\nu}\}=
\{\bar{q}_{\mu},\bar{q}^{\dagger}_{\nu}\}=\delta_{\mu \nu}, 
\nonumber\\
&&
\{Q_{\mu}, Q^{\dagger}_{\nu}\}=
\{\bar{Q}_{\mu},\bar{Q}^{\dagger}_{\nu}\}=\delta_{\mu \nu}, 
\label{aqcom}
\end{eqnarray}
all other (anti)commutators are zero. The composite operators $N$, $S$ and $G$ in
Eqs.~(\ref{m-meson2})-(\ref{qq-gg}) have non-canonical commutators
\begin{eqnarray}
\!\!\!\!\!\!
&&\lbrack N_{\alpha}, N^{\dagger}_{\beta} \rbrack=\delta_{\alpha \beta} -{\cal N}_{\alpha \beta}
\,\,\,;\,\,\, 
\lbrack S_{\alpha}, S^{\dagger}_{\beta} \rbrack=\delta_{\alpha \beta} -{\cal S}_{\alpha \beta}
\nonumber\\&&
\lbrack G_{\alpha},G_{\beta}^{\dagger}\rbrack=\delta_{\alpha\beta}+{\cal G}_{\alpha\beta}
\label{QGcom}
\end{eqnarray}
where
\begin{eqnarray}
{\cal N}_{\alpha \beta}&=& \varphi_{\alpha}^{*{\mu \nu }}
\varphi_{\beta}^{\mu \sigma }\bar{q}^{\dagger}_{\sigma}\bar{q}_{\nu}
+ \varphi_{\alpha}^{*{\mu \nu }}
\varphi_{\beta}^{\rho \nu}q^{\dagger}_{\rho}q_{\mu}
\nonumber\\
{\cal S}_{\alpha \beta}&=& \Upsilon_{\alpha}^{*{\mu \nu }}
\Upsilon_{\beta}^{\mu \sigma }\bar{Q}^{\dagger}_{\sigma}\bar{Q}_{\nu}
+ 
\Upsilon_{\alpha}^{*{\mu \nu }}
\Upsilon_{\beta}^{\rho \nu}Q^{\dagger}_{\rho}Q_{\mu}
\nonumber\\
 {\cal G}_{\alpha\beta}&=&2\Psi_{\alpha}^{\ast\mu\gamma}\Psi_{\beta}^{\gamma \rho}a_{\rho}^{\dagger}a_{\mu}. 
\label{mg}
\end{eqnarray}
The presence of ${\cal N}_{\alpha\beta}$, ${\cal S}_{\alpha\beta}$ and ${\cal G}_{\alpha\beta}$ in Eq.~(\ref{QGcom})
reflects the composite nature of the meson. In the Fock-Tani formalism, the physical particles are replaced
by ``ideal particles'' and $F^\dag$ must be replaced by a new creation op\-era\-tor
\begin{eqnarray}
f^{\dag}_{\alpha}=c_1\,n_{\alpha}^{\dagger}+c_2\,s_{\alpha}^{\dagger} +c_3\,g_{\alpha}^{\dagger}\,,
\label{ideal-f}
\end{eqnarray}
where canonical relations are satisfied:
\begin{eqnarray*}
%\hspace{-.25cm} 
&&[n_{\alpha},n_{\beta}^{\dagger}]  =\delta_{\alpha\beta}
\,\,\,\,\,;\,\,\,\, 
[s_{\alpha},s_{\beta}^{\dagger}]  =\delta_{\alpha\beta}
%\nonumber\\
%&&
\,\,\,\,\,;\,\,\,\, 
[g_{\alpha},g_{\beta}^{\dagger}]  =\delta_{\alpha\beta}.
\label{q-g-com}
\end{eqnarray*}
To obtain Eq.~(\ref{ideal-f}), the Fock-Tani formalism requires the definition of a unitary transformation
$U$ that maps the composite state onto an ideal state, {\it i.e.},
\begin{eqnarray}
&& U^{-1}\,N_{\alpha}^{\dagger}\,|0\rangle =n_{\alpha}^{\dagger}\,|0\rangle
\,\,\,\,\,;\,\,\,\, 
U^{-1}\,S_{\alpha}^{\dagger}\,|0\rangle =s_{\alpha}^{\dagger}\,|0\rangle
\nonumber\\&&
U^{-1}\,G_{\alpha}^{\dagger}\,|0\rangle=
\,g_{\alpha}^{\dagger}\,|0\rangle.
\label{U-QG}
\end{eqnarray}
The mapping (\ref{U-QG}) is achieved when one writes $U=\exp({t\,{\cal F}})$ and the parameter $t$
assumes the value of $-\pi/2$. The operator ${\cal F}$ is the generator of the transformation given by
\begin{eqnarray}
{\cal F}={\cal F}_N+{\cal F}_S+{\cal F}_G\,,
\end{eqnarray}
where ${\cal F}_N$, ${\cal F}_S$ and ${\cal F}_G$ are defined as
\begin{eqnarray}
%\!\!\!\!\!\!\!\!\!\!\!\!
{\cal F}_N&=& n_{\alpha}^{\dagger}\,\tilde{N}_{\alpha}  - \tilde{N}_{\alpha}^{\dagger}\,n_{\alpha}
\,\,\,;\,\,
{\cal F}_S= s_{\alpha}^{\dagger}\,\tilde{S}_{\alpha}  - \tilde{S}_{\alpha}^{\dagger}\,s_{\alpha}
\nonumber\\
{\cal F}_G &=& g_{\alpha}^{\dagger}\,\tilde{G}_{\alpha}  - \tilde{G}_{\alpha}^{\dagger}\,g_{\alpha}.
  \label{F}
\end{eqnarray}
The operators $\tilde{N}$, $\tilde{S}$ and $\tilde{G}$ are expansions in powers of the wave-function,
with the following conditions
\begin{eqnarray}
&&\lbrack\tilde{N}_{\alpha},\tilde{N}^{\dagger}_{\beta}\rbrack = \delta_{\alpha\beta}+
{\cal O} (\varphi^{n+1})
%\,\,\,;\,\,\,
\nonumber\\
&&\lbrack\tilde{S}_{\alpha},\tilde{S}^{\dagger}_{\beta}\rbrack = \delta_{\alpha\beta}+
{\cal O} (\Upsilon^{n+1})
%\,\,\,;\,\,\,
\nonumber\\
&&\lbrack\tilde{G}_{\alpha},\tilde{G}^{\dagger}_{\beta}\rbrack = \delta_{\alpha\beta}+
{\cal O} (\Psi^{n+1})\,. 
\label{comO}
\end{eqnarray}
It is easy to see from Eq.~(\ref{F}) that ${\cal F}^{\dag}=-{\cal F}$, which ensures that $U$ is unitary.
The operators $\tilde{N}_{\alpha}$, $\tilde{S}_{\alpha}$ and $\tilde{G}_{\alpha}$ are determined up
to a specific order $n$ consistent with Eq.~(\ref{comO}). The examples studied in \cite{annals}
required the determination in the quark sector, for example, $\tilde{N}^{(i)}_{\alpha}$, up to
order 3 as shown below
\begin{eqnarray}
\tilde{N}_{\alpha}^{(0)}&=&N_{\alpha}  \hspace{.5cm};  \hspace{.5cm}
\tilde{N}_{\alpha}^{(1)}= 0  
%\nonumber\\
\hspace{.5cm};  \hspace{.5cm}
\tilde{S}_{\alpha}^{(2)} =  \frac{1}{2}{\cal N} _{\alpha  \beta}\,N_{\beta}
 % \hspace{.1cm};  \hspace{.1cm}
\nonumber\\
\tilde{N}_{\alpha}^{(3)} &=&\frac{1}{2}N^{\dagger}_{\beta}\,\,
[N_{\alpha},{\cal N}_{\alpha\beta}]\,\,N_{\gamma}.
\label{mes_gen2} 
\end{eqnarray}
In the glueball mapping, the transformed operators are similar to the meson's and details are
described in \cite{mario}. By the former discussion it is trivial to show that the mixed meson
$F^{\dag}$ is also mapped onto the ideal sector
\begin{eqnarray}
|\alpha\,)&=&   U^{-1}\,F_{\alpha}^{\dagger}\,|0\rangle=
f_{\alpha}^{\dagger}\,| 0\rangle.
\label{U}
\end{eqnarray}

Applying the Fock-Tani formalism to a microscopic Hamiltonian $H$ gives rise to an effective
interaction ${\cal H}_{FT}$,
\begin{eqnarray}
  {\cal H}_{FT} = U^{-1} H U\,.
\label{hft}
\end{eqnarray}
To find this  Hamiltonian we have to calculate the transformed operators for quarks, antiquarks
and gluons by a technique known as {\it the equation of motion technique} described in references
\cite{annals,prd08}. The Fock-Tani formalism is a general theoretical framework, limited by the
choice of the microscopic Hamiltonian, which in many cases is defined by phenomenological
assumptions. In the quark sector, for example, a very successful approach is a pair production
Hamiltonian that regards the decay of an initial state meson in the presence of a $q\bar{q}$ pair 
created from the vacuum. The pair is obtained from the non-relativistic limit of the  interaction
Hamiltonian $H_{q\bar{q}}$ involving Dirac quark fields \cite{prd08,barnes}
\begin{eqnarray} 
 H_{q\bar{q}}&=&g_{q\bar{q}}\int d\vec{x}\,\bar{\psi}(\vec{x})\,\psi(\vec{x}).
\label{3p0}
\end{eqnarray}
The procedure to obtain an effective bound-state corrected Hamiltonian from the $^{3}P_{0}$ interaction
(\ref{3p0}) after the Fock-Tani mapping was  described in \cite{prd08} and called the
{\it $C^{\,3}P_0$ Hamiltonian},
\begin{eqnarray}
\!\!\!\!
{\cal H}^{\rm C3P0} &=&
-\Phi^{\ast\rho\xi}_{\alpha}
\Phi^{\ast\lambda\tau}_{\beta}\Phi^{\mu\nu}_{\gamma}
%\left(\bar{a}_1\,\varphi^{\mu\nu}_{\gamma}+  \bar{a}_2\,\Ups^{\mu\nu}_{\gamma}\, \right) 
\,V\, m^{\dag}_{\alpha} m^{\dag}_{\beta} m_{\gamma}.
\label{c3p0}
\end{eqnarray} 
where in Eq.~(\ref{c3p0}) $m$ is an generic ideal meson ($n$ or $s$); the potential $V $ is a
condensed notation for 
\begin{eqnarray}
\!\!\!\!\!\!\!\!\!\!\!\!\!\!\!\!\!\!\!\!\!\!
 V&=&V_{\mu\nu}
\left[\delta_{\mu\lambda}
\delta_{\nu\xi}
\delta_{\omega\rho}
\delta_{\sigma\tau}
-
\frac{1}{2}
\delta_{\sigma\xi}\,\delta_{\lambda\omega}\,\,
\Delta(\rho\tau;\mu\nu)
\right.
\nonumber\\
&&
+
\frac{1}{4}
\delta_{\sigma\xi}\,\delta_{\lambda\mu}\,\,
\Delta(\rho\tau;\omega\nu)
+
\left.
\frac{1}{4}
\delta_{\xi\nu}\,\delta_{\lambda\omega}\,\,
\Delta(\rho\tau;\mu\sigma)
\right]
\label{vc3p0}
\end{eqnarray}
while the pair creation potential $V_{\mu\nu}$ is given by
\begin{eqnarray} 
V_{\mu\nu}\equiv 2\,m_{q}\, \gamma\,
\,\delta(\vec{p}_{\mu}+\vec{p}_{\nu})\, \bar{u}_{s_{\mu}f_{\mu}c_{\mu}  }
(\vec{p}_{\mu}) \, v_{s_{\nu} f_{\nu}c_{\nu}  }(\vec{p}_{\nu}), 
\label{vmn} 
\end{eqnarray}
with $g_{q\bar{q}}=2\,m_{q}\, \gamma$. It should be noted that since Eq.~(\ref{3p0}) is meant to be
taken in the non-relativistic limit, Eq.~(\ref{vmn}) should be as well. The bound-state kernel
$\Delta$ in (\ref{vc3p0}) is defined as 
\begin{eqnarray}
\Delta(\rho\tau;\mu\sigma)=\Phi_{\alpha}^{ \rho\tau}\Phi_{\alpha}^{\ast \mu\sigma} .
\label{kernel}
\end{eqnarray}
The physical implications of (\ref{vc3p0}) and (\ref{kernel}) were discussed in detail in reference
\cite{prd08}.

The glue sector is not covered by the former discussion and a consistent map implies that the
microscopic Hamiltonian must contain the quark-gluon vertex. In different approaches, the glueball 
is pictured as a non-relativistic bound-state of two constituent gluons, where decay proceeds by the
conversion of these gluons into $q\bar{q}$ pairs, again the  interaction vertex is the crucial ingredient.
An elementary interaction vertex used in \cite{faessler} is given to lowest order in the
non-relativistic limit by
\begin{equation}
V^{(1)}_{(a_i  q_l\bar{q}_k)}=g_{G}\, 
(\vec{\sigma}_{(lk)}\cdot\vec{\epsilon}_{i})~{\mathbf{1}}_F^{(lk)}~
\left(\frac{1}{2}\sum_{b=1}^8 \lambda^b_{(lk)}A^b_i\right)
\label{vertex}
\end{equation}
where the vertex in Eq.~(\ref{vertex}) is multiplied by an overall momentum conservation
$\delta(\vec k_i-\vec p_l-\vec p_k)$, with the internal momenta $\vec k_i$ for gluon $i=1,2$ and
$\vec p_{l(k)}$ for the  quarks with label $l=1,3$ ($k=2,4$). The identity operator 
${\mathbf{1}}^{(lk)}_F$ projects onto a flavor singlet state of the created $q\bar{q}$ pair ($lk$).
The last term in Eq.~(\ref{vertex}) is the color part of the interaction vertex with the
Gell-Mann matrices $\lambda^b$ acting in color space of ($lk$). The color octet wave-function of the
gluon $i$ with polarization vector $\vec{\epsilon}_i$ is denoted by $A^b_i$. An effective second
order amplitude with a four quark-antiquark and two gluon operator structure of the type
\begin{eqnarray}   
V^{(2)}_{(a_i  q_l\bar{q}_k)}\sim \left(q^\dag \bar{q}^\dag\,a\right)\left( q^\dag \bar{q}^\dag \,a\,\right)
\label{vertex2}
\end{eqnarray}
can be obtained from Eq.~(\ref{vertex}). This is a required assumption that guarantees, in the glue sector
of $f_0$, that after the Fock-Tani transformation, the ideal glueball $g$ decays to {\it ideal mesons}
$m$ \cite{mario}. The effective Hamiltonian ${\cal H}_{G}$ is obtained from Eq.~(\ref{vertex2}), in lowest
order,
\begin{eqnarray}
  {\cal H}_{G} &=& U^{-1}\,V^{(2)}_{(a_i  q_l\bar{q}_k)}\,U
\nonumber\\
&\approx&
V_I \,\, 
\Phi_\beta^{\ast\mu\rho} \,
\Phi_\delta^{\ast\nu\eta}\,        
\Psi_\alpha^{\tau\xi} \,
m_\beta^\dag \, m_\delta^\dag \,g_\alpha\,
\label{eq323}
\end{eqnarray}
with
\begin{eqnarray}
  V_I &\equiv & \bar{\delta}^2  \frac{\alpha_s}{
\, 8\sqrt{2}\pi^{2}} 
\frac{\lambda^{b_\tau} \lambda^{b_\xi}}{  \sqrt{  \omega_{\vec{p}_\tau}   \omega_{\vec{p}_\xi}}}
\,
 \Pi_{\mu\,\nu}({\cal P}_\tau)\,\Pi_{\eta\,\rho}({\cal P}_\xi)
\end{eqnarray}
where  $\alpha_s=g_G^2/(4\pi)$, $ \Pi_{\mu\,\nu}({\cal P}_\tau)=\vec{\sigma}_{\mu\nu} \cdot
\vec{\epsilon}({\cal P}_\tau)$, ${\cal P}$ is the gluon's polarization vector and $ \bar{\delta}^2= 
\delta(\vec{p}_\mu + \vec{p}_\nu - \vec{p}_\tau)\delta(\vec{p}_\sigma + \vec{p}_\rho 
- \vec{p}_\xi)$.

The wave-function of the $f_0$ meson has quark-an\-ti\-quark and glueball components. The quark-antiquark
sector wave-function $\varphi$ (and/or $\Upsilon$) is written as the following product
\begin{eqnarray}
\varphi_{\alpha}^{\mu\nu}=
\chi_{S_{\alpha}}^{s_{\mu}s_{\nu}}\,
{\cal C}^{c_{\mu}c_{\nu}}\,
\xi^{f_{\mu}f_{\nu}}\,
\varphi_{\vec{P}_{\alpha} }^{\,\vec{p}_{\mu}\vec{p}_{\nu}},
\label{wf-quark1}
\end{eqnarray}
$\chi$ is the spin contribution ($S_{\alpha}$ is the meson's spin); ${\cal C}$ is the color 
component; $\xi$ is the flavor part and the spatial wave-function is
\begin{eqnarray}
\varphi_{\vec{P}_{\alpha}}^{\vec{p}_{\mu}\vec{p}_{\nu}}=
\delta^{(3)}(\vec{P}_{\alpha}-\vec{p}_{\mu}-\vec{p}_{\nu})\,
{\left(\frac{1}{\pi
\beta_{q}^2}\right)}^{\frac{3}{4}}e^{-\frac{1}{8\beta_{q}^{2}}{
\left(\vec{p}_{\mu}-\vec{p}_{\nu}\right)}^{2}}.
\label{wf-quark2}
\end{eqnarray}
The glueball wave-function $\Psi$ has a similar structure to (\ref{wf-quark1}) and (\ref{wf-quark2})
with the parameter $\beta_q$ replaced by $\beta_g$ and with the flavor part absent in
(\ref{wf-quark1}). In our example, the final state wave-function $\Phi$ is for pions, where again
the form written in Eqs.~(\ref{wf-quark1}) and (\ref{wf-quark2}) shall be used with the following
substitution $\beta_q\to\beta_{\pi}$.

To determine the decay rate, we define the initial and final states by $|i\rangle=f^{\dag}_{\alpha}
|0\rangle$ and $|f\rangle=m^{\dag}_{\beta} m^{\dag}_{\gamma} |0\rangle$. The matrix element between these
states is 
\begin{eqnarray}
\langle f \mid ({\cal H}^{C3P0}+ {\cal H}_{G}) \mid  i \rangle= \delta (\vec{p}_{\alpha}-\vec{p}_{\beta}-\vec{p}_{\gamma})h_{fi}, 
\end{eqnarray}
where the decay amplitude $h_{fi}$ can be written as
\begin{eqnarray}
h_{fi}=c_1\,h^{n\bar{n}}_{fi}+c_2\,h^{s\bar{s}}_{fi}+c_3\,h^{g}_{fi},
\label{hfi}
\end{eqnarray}
where in Eqs.~(\ref{ideal-f}) and (\ref{hfi}) the following was considered: $c_1=b_1 a_1$, $c_2=b_1 a_2$ and
$c_3=b_2 $. This assumption is consistent with Eq.~(\ref{f0}) and the normalization condition on $c_i$.

\begin{figure}[ht]
\includegraphics[scale=0.6]{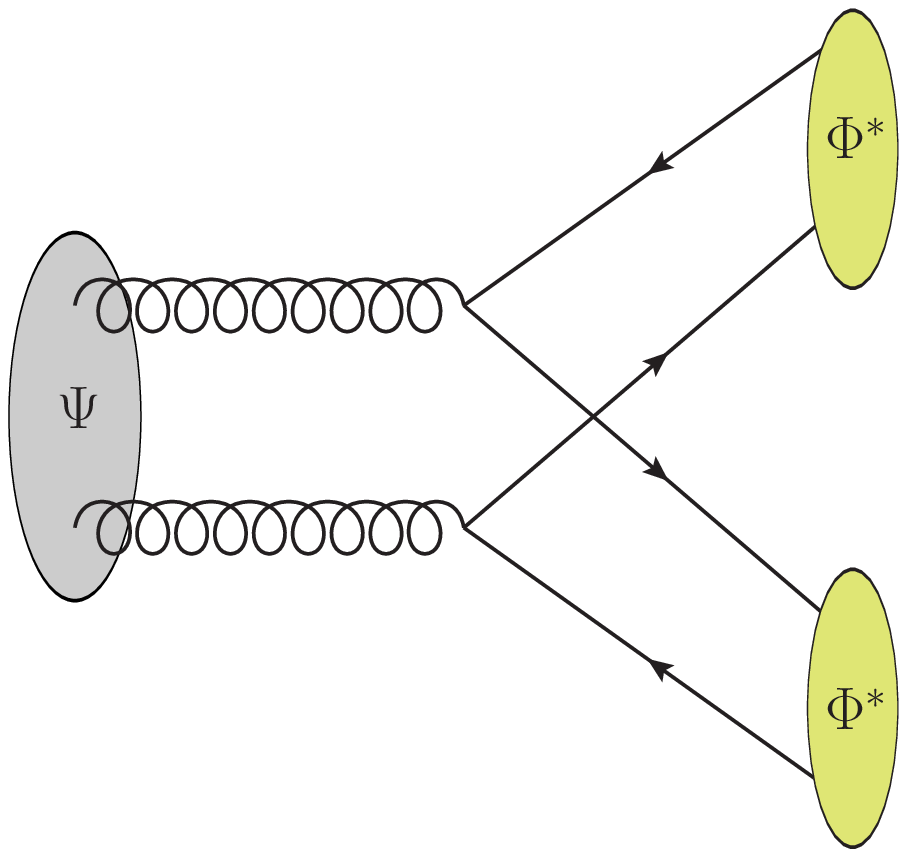}
\includegraphics[scale=0.53]{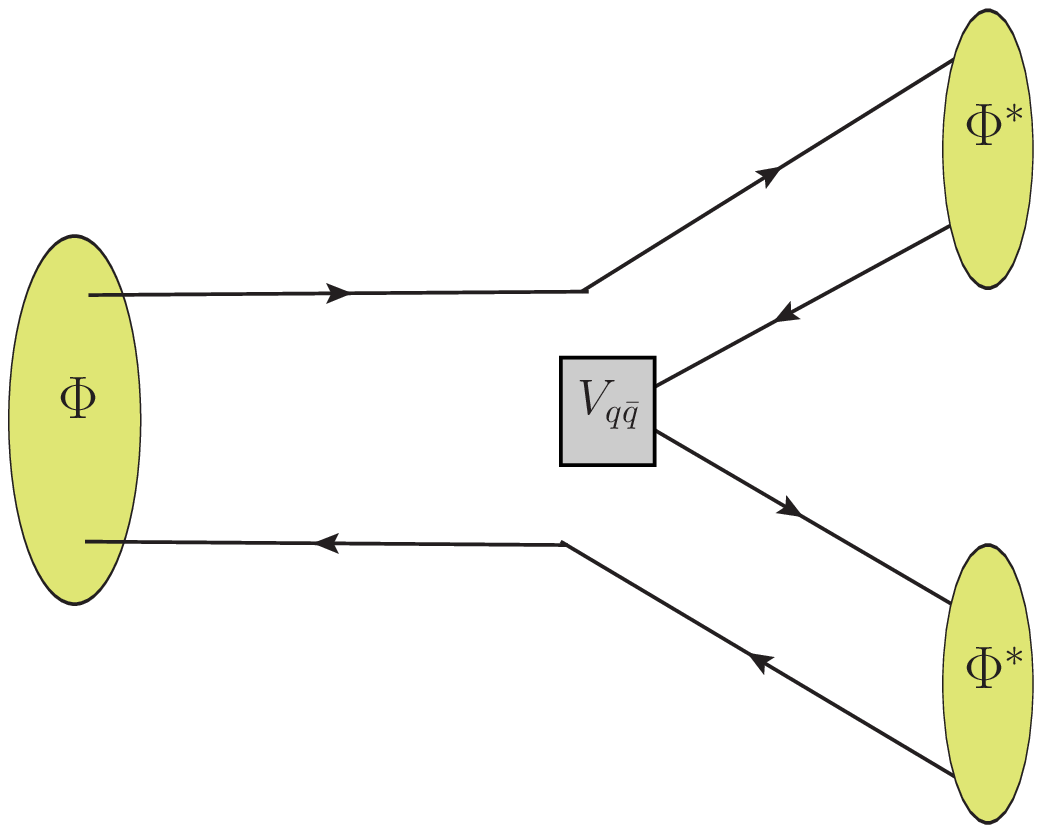}
\caption{Fock-Tani amplitude is a mixture of a glueball decay into mesons (upper) and the meson
decay (lower).}
\end{figure}

\begin{figure}[ht]
\includegraphics[scale=0.35]{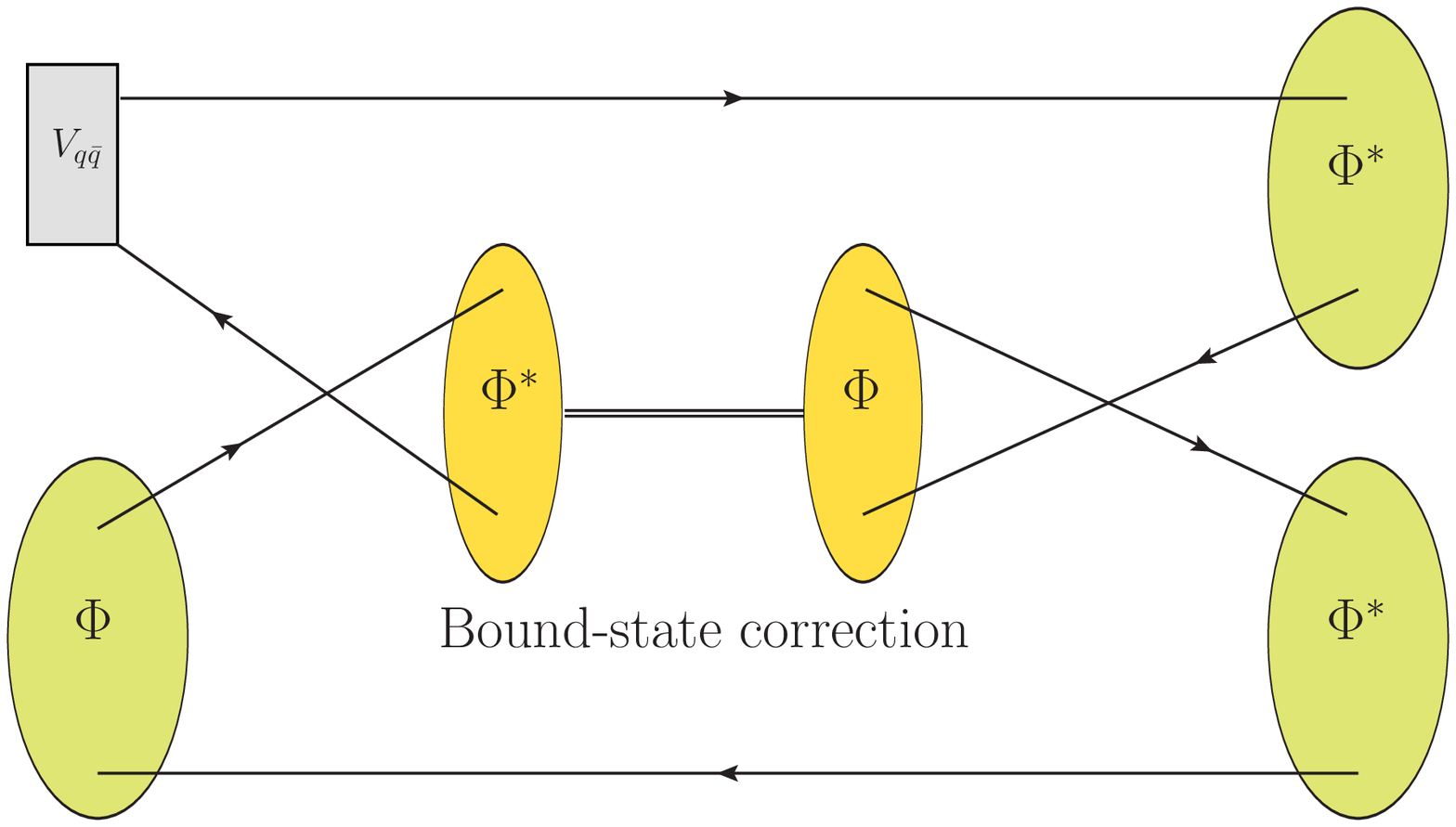}
\includegraphics[scale=0.35]{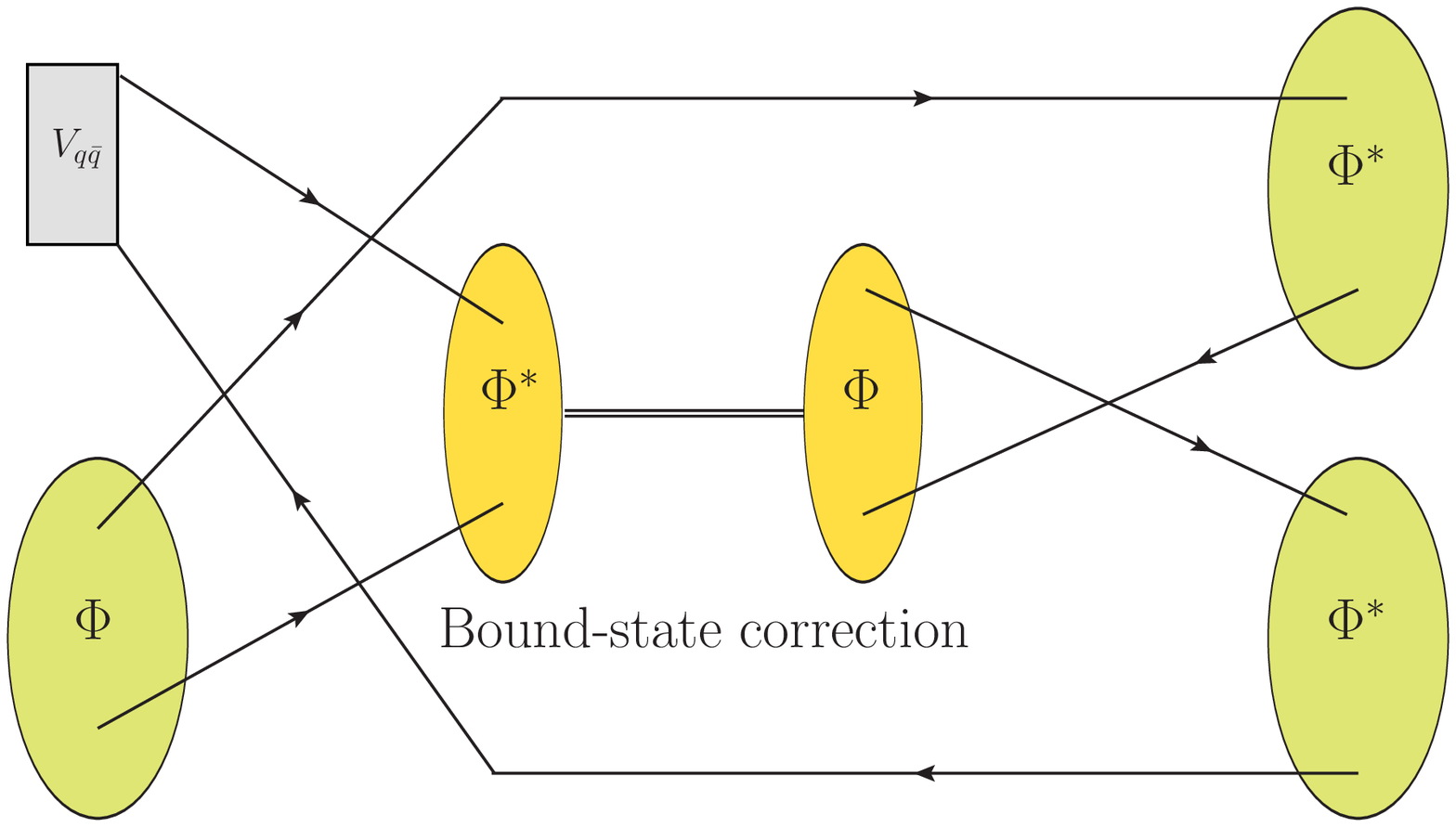}
\caption{Fock-Tani bound-state correction amplitudes to the $q\bar{q}$ pair creation. There is no
correction to the glueball amplitude.}
\end{figure}

\section{Numerical example}
 
In our example we shall study the following decay channel $f_0(1500)\to 2 \pi$. The amplitudes
obtained are
\begin{eqnarray}
h^{n\bar{n}}_{fi}&=&
  \gamma\, 
\left[b_1(p)\,
e_1(p)
-b_2(p)\,
e_2(p)
\right] \,Y_{00}
\nonumber\\
h^{s\bar{s}}_{fi}&=&0
\nonumber\\
h^{g}_{fi}&=&  \,\alpha_s \, b_3\,Y_{00}
\label{hfi-tot}
\end{eqnarray}
where $Y_{00}$ is the spherical harmonic and
\begin{eqnarray}
b_1(p)&=&\frac{32}{\pi ^{1/4}} \sqrt{\frac{2}{3}} 
\beta_{q}^{5/2} \frac{3 \beta_{\pi}^{2} (2 \beta_q^2 +\beta_{\pi}^2) - (\beta_q^2+\beta_{\pi}^2 )\,p^2
}{
2\left(2 {\beta_q}^2+{\beta_\pi}^2\right)^{7/2}    }
\nonumber\\
b_2(p)&=&\frac{32}{\pi ^{1/4}} \sqrt{\frac{2}{3}} 
\beta_{q}^{5/2} \frac{3\beta_{\pi}^{2}( 4\beta_q^2 + 3 \beta_{\pi}^2 )-4 (\beta_q^2+ \beta_{\pi}^2)\,p^2
}{
3 \left(    4   \beta_q^2   + 3  \beta_{\pi}^2  \right)^{7/2}
    }
\nonumber\\
b_3&=&\frac{4 \sqrt{2} }{9\,\pi^{9/4} } \beta_g^{1/2}\,  \beta_{\pi}^2 
  \frac{ (  3 \beta_g^2         -2\beta_{\pi}^2       )}{   (\beta_g^2 + 2\beta_{\pi}^2)^2 \, }\,\,
\nonumber\\
e_1(p)&=&\exp\left(-\frac{p^2}{8 {\beta_q}^2+4 {\beta_\pi}^2}\right)
\nonumber\\
e_2(p)&=&\exp\left(-\frac{\left(4  { \beta_q}^2+5  { \beta_{\pi}}^2\right)  {p}^2}
{4    { \beta_{\pi}}^2 \left(4  { \beta_q}^2+3  { \beta_{\pi}}^2\right)}\right)
\label{b123}
\end{eqnarray}
The second term  in $h^{n\bar{n}}_{fi}$ is the bound-state correction for the light quark sector.  
If one sets $b_2(p)=0$, the bound state correction is turned off.

The $h_{fi}$ decay amplitude can be combined with a relativistic phase space to give the
differential decay rate \cite{barnes}
\begin{eqnarray}
\frac{d\Gamma_{\alpha\rightarrow \beta\gamma}}{d\Omega}=2\pi\, \frac{PE_{\beta}E_{\gamma} }{ M_{\alpha}}
\,\,|h_{fi}|^2=2\pi\, \frac{PE_{\pi}^2 }{ M_{f_0}}
\,\,|h_{fi}|^2,
\end{eqnarray}
which after integration in the solid angle $\Omega$ gives rise to the decay rate, a  usual
choice for the meson momenta is made: $\vec{P}_{f_0}=0$  ($P=|\vec{P}_\pi|$). The experimental
value for this decay channel is $\Gamma_{\rm exp}=38.04\,\pm {2.51}$ MeV \cite{pdg}. The meson
masses assumed in the numerical calculation  have standard values of $M_{\pi}=138$ MeV and
$M_{f_0}=1505$ MeV. There are two other sets of parameters, the first is the pair of coupling
constants $\gamma$ and $\alpha_s$; the second are  three wave-function widths $\beta_{\pi}$,
$\beta_q$ and $\beta_g$. The value of $q\bar{q}$ coupling can be extracted from light meson,
$\gamma= 0.5$ \cite{barnes}. The parameter $\alpha_s$ is the quark-gluon coupling, which is
assumed to be fixed at the usual value $0.6 $. The quark sector widths are in the range of 0.3
- 0.4 GeV \cite{barnes}. For the pion, we shall fix $\beta_{\pi}=0.4$ GeV. The parameter
$\beta_q$ is from the quark sector of $f_0(1500)$ and is chosen close, but slightly smaller
than the pion's value $\beta_q = 0.3$ GeV. This leaves only one actually free parameter, the
glueball's width $\beta_g$, that should be adjusted. The best fit of  $\Gamma$ to the range of
the experimental value results in $\beta_g$ the order of 1.6 GeV. The mixing parameters $c_i$
used are from Ref. \cite{faessler} where three models (A, B and C) are studied in detail. A
comparison is shown in table (\ref{tab1}).
 
In conclusion, we have showed that the Fock-Tani formalism applied to meson decay with mixing
is a promising approach and it exhibits bound-state corrections to the decay amplitude. This
extends the OZI-allowed decays beyond the ordinary $^{3}P_{0}$ approach. Even though the
choice of the microscopic Hamiltonians for the quark and gluon sectors is very simple, the
essence of the procedure is clear and introduces a difference ranging from 7\% to 11\% in the
decay rates. In future calculations an approach based on a Hamiltonian formulation, for
example, QCD in the Coulomb gauge \cite{swanson1,swanson2,swanson3}, can replace some of the
unknown parameters by fundamental quantities, extracted from QCD.

\begin{table}
\caption{Comparison of the  decay rates in the $^{3}P_{0}$  and C$^{3}P_{0}$ models  }
\begin{tabular}{cccc|ccc}
  \hline \hline
      &       &       &       &       &    $\Gamma$\,\, (MeV) & \\  
      %&       &       &       &       &                    \\
Model & $c_1$ & $c_2$ & $c_3$ & $^{3}P_{0}$  &                       &  C$^{3}P_{0}$ 
\\
\hline
&&&&&&
\\
A     & 0.43  & -0.61 & 0.61  & 37.6  &              &   41.2    \\
B     & 0.40  & -0.90 & 0.19  &  19.8 &              &    22.3         \\
C     & 0.31  & -0.58 & 0.75  & 30.1  &              &   32.5           \\
\hline \hline
\label{tab1}
\end{tabular}
\end{table}

\section*{Acknowledgments}
This research was supported by Conselho Nacional de Desenvolvimento Cient\'{\i}fico
e Tecnol\'ogico (CNPq), Universidade Federal do Rio Grande do Sul (UFRGS) and
Universidade Federal de Pelotas (UFPel).

%% The Appendices part is started with the command \appendix;
%% appendix sections are then done as normal sections
%% \appendix

%% \section{}
%% \label{}

\end{document}